\documentclass[letterpaper, 10 pt, conference]{ieeeconf}  

\IEEEoverridecommandlockouts                              

\usepackage{cite}
\usepackage{amsmath,amssymb,amsfonts}
\usepackage{graphicx}
\usepackage{textcomp}
\usepackage{amsmath} 
\usepackage{amssymb}  

\usepackage{algorithm}
\usepackage{algorithmicx}
\usepackage[noend]{algpseudocode}
\usepackage{balance}
\renewcommand{\baselinestretch}{.985}

\title{\LARGE \bf
Symptom-Driven Personalized Proton Pump Inhibitors Therapy Using Bayesian Neural Networks and Model Predictive Control
}

\author{Yutong Li and Ilya Kolmanovsky$^{1}$
\thanks{$^{1}$Yutong Li and Ilya Kolmanovsky are with the Department of Aerospace Engineering, University of Michigan at Ann Arbor, Ann Arbor, MI 48105 USA. Emails:
{\tt\small wilson420813@gmail.com} and {\tt\small ilya@umich.edu}}%
}

\begin{document}

\maketitle
\thispagestyle{empty}
\pagestyle{empty}

\begin{abstract}

Proton Pump Inhibitors (PPIs) are the standard of care for gastric acid disorders but carry significant risks when administered chronically at high doses. Precise long-term control of gastric acidity is challenged by the impracticality of invasive gastric acid monitoring beyond 72 hours and wide inter-patient variability. We propose a noninvasive, symptom-based framework that tailors PPI dosing solely on patient-reported reflux and digestive symptom patterns. A Bayesian Neural Network prediction model learns to predict patient symptoms and quantifies its uncertainty from historical symptom scores, meal, and PPIs intake data. These probabilistic forecasts feed a chance-constrained Model Predictive Control (MPC) algorithm that dynamically computes future PPI doses to minimize drug usage while enforcing acid suppression with high confidence—without any direct acid measurement. In silico studies over diverse dietary schedules and virtual patient profiles demonstrate that our learning-augmented MPC reduces total PPI consumption by 65\% compared to standard fixed regimens, while maintaining acid suppression with at least 95\%  probability. The proposed approach offers a practical path to personalized PPI therapy, minimizing treatment burden and overdose risk without invasive sensors.

\end{abstract}

\section{Introduction}\label{sec:intro}

Gastric acid–related disorders affect over 20 million people in the United States and contributed to 770000 gastric cancer deaths worldwide in 2020 \cite{morgan2022current}.  Proton Pump Inhibitors (PPIs) are the standard therapy for these conditions, effectively suppressing acid secretion by raising gastric pH \cite{reimer2013safety,huang2001pharmacological}.  However, chronic or excessive PPI use can lead to serious side effects—osteoporosis, nausea, and rebound hyperacidity on abrupt cessation—and optimal, patient‐specific dosing remains an open challenge \cite{vaezi2017complications}.

A common personalization strategy relies on clinical trials with direct pH monitoring to titrate doses \cite{shin2013pharmacokinetics,shin2008pharmacology,lundell2015systematic}.  Standard 24–48 hour catheter‐based pH probes deliver accurate acidity profiles but cause nasal and pharyngeal discomfort and limit patient activity \cite{gyawali2024updates}.  Wireless capsule systems (e.g.\ Bravo) improve tolerability but remain approved only for short‐term use (up to 96 hours) and can still provoke chest pain in a subset of patients \cite{gyawali2024updates}. 

An alternative is to leverage ODE models of gastric physiology to design dosing schedules without continuous measurement.  Early two‐compartment models captured neural and hormonal control of acid secretion \cite{de1993gastric,livcko1992dual}, and later extensions incorporated detailed gastrin–histamine–somatostatin kinetics \cite{joseph2003model,sud2004predicting}.  While these models offer mechanistic insight, they contain dozens of rate and Michaelis–Menten constants that must be calibrated from typically only 1–3 days of pH or secretion data.  Structural identifiability and parameter estimation studies demonstrate that many parameters cannot be uniquely inferred from such limited data, leading to large uncertainties and poor long‐term predictive accuracy \cite{van2024navigating,li2023scheduling}.

To address both the burden of long‐term invasive monitoring and the parameter‐identifiability limitations of mechanistic models, we propose a fully noninvasive, symptom‐driven PPI dosing framework.  Our method relies solely on patient‐reported reflux and digestive discomfort scores—validated proxies for gastric health \cite{shaw2008reflux}—collected alongside meal and dosing logs.  A sequence‐to‐sequence Bayesian Neural Network (BNN) is trained to predict these symptom trajectories and to quantify its own uncertainty using only historical symptom, meal, and dose data.  These probabilistic forecasts are then fed into a chance‐constrained Model Predictive Control (MPC) algorithm, which dynamically computes the minimal PPI dose adjustments needed to maintain symptoms below clinician‐defined thresholds with high confidence—entirely without direct pH measurements.

The main contributions of this paper are:
\begin{enumerate}
  \item A noninvasive, symptom‐based PPI dosing framework that eliminates the need for invasive acid monitoring.
  \item A sequence‐to‐sequence Bayesian Neural Network (BNN) that predicts future reflux and digestive symptoms, complete with uncertainty estimates.
  \item A chance‐constrained MPC algorithm that leverages the BNN forecasts to compute minimal PPI schedules satisfying symptom constraints with high probability.
  \item In silico validation demonstrating personalized dosing schedules that reduce PPI consumption by over 65\% compared to standard fixed regimens, while maintaining symptom control.
\end{enumerate}

\section{System Architecture for Symptom-driven Personalized PPI Dosing}
\label{sec:system_architecture}

The overall architecture of our learning‐based, symptom‐driven PPI dosing system is shown in Fig.~\ref{fig:sys_arch}.  At its core, personalized PPI scheduling relies exclusively on patient‐reported symptom scores, thereby eliminating invasive gastric acid monitoring—a key obstacle to precise, long‐term PPIs dosing. Historical symptom trajectories and corresponding PPI dose records are used to train, offline, a Bayesian Neural Network (BNN) that both forecasts future symptom profiles and quantifies predictive uncertainty. This BN is then embedded within a chance‐constrained Model Predictive Control (MPC) framework, which computes the minimal PPI dosing schedule necessary to satisfy prescribed symptom constraints with high probability.  The key modules of the proposed system are detailed below:

\begin{figure*}[t]
  \centering
  \includegraphics[width=0.6\textwidth]{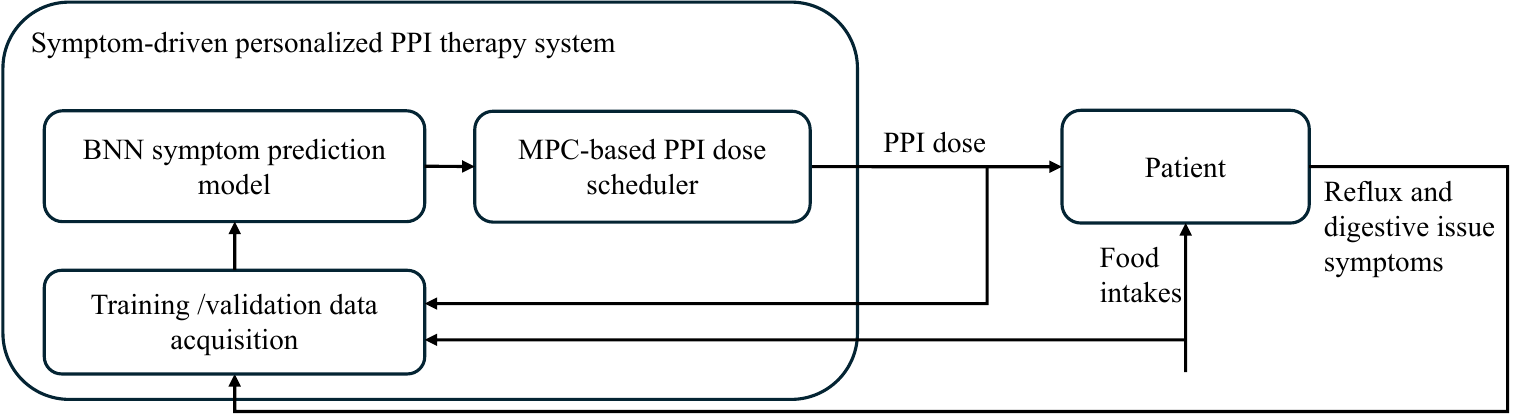}
  \caption{End-to-end architecture for learning‐based, symptom‐driven PPI dosing.  (1) \emph{Data Acquisition} collects patient‐reported symptom scores and meal/dosing logs via mobile or clinical interfaces.  (2) A \emph{Bayesian Neural Network (BNN)} is trained on historical and real‐time data to forecast future symptom trajectories with quantified uncertainty.  (3) A \emph{Chance‐Constrained MPC} module utilizeds the BNN forecasts and multiple meal scenarios to compute a minimal PPI dose schedule that guarantees symptom targets with high probability.  (4) The recommended dosing is recommended to the patient (or clinician) and the loop repeats at each dosing interval.}
  \label{fig:sys_arch}
\end{figure*}

\paragraph{Data Acquisition and Symptom Encoding}
Patients report two noninvasive symptom scores—\emph{reflux severity} and \emph{digestive discomfort}—on a 1–10 scale, together with timestamps for meals and any administered PPI doses. Reflux severity reflects episodes of excessive gastric acidity, whereas digestive discomfort captures symptoms arising from insufficient acid; the combination provides a proxy for acid homeostasis. Data may be entered via smartphone app, web portal, or electronic health record, eliminating the need for invasive gastric acid probes. Raw reports are then time‐aligned and noise‐filtered, and organized into a sliding window of recent history for both offline BNN training and online inference.

\paragraph{Learning‐Based Surrogate Model}
A Bayesian Neural Network (BNN) is trained to map recent symptom–meal–dose histories to future symptom trajectories. Initially, we pretrain a foundation BNN offline on aggregated patient data over a fixed time window. To capture individual variability, this foundation model is then rapidly fine‐tuned on each patient’s own historical records, enabling data‐efficient personalization. At runtime, the BNN delivers both a point forecast of symptom scores over the prediction horizon and an uncertainty estimate via Monte Carlo dropout, supporting risk‐aware dosing decisions making.

\paragraph{Chance-Constrained PPI dose scheduling}
Leveraging the BNN’s probabilistic forecasts, a Model Predictive Control (MPC) optimizer searches for the minimal sequence of daily PPI dose that keeps predicted symptom scores below a clinician-set threshold with high probability (e.g.\ 95 \%). To guard against uncertain meal patterns, each candidate dosing plan is evaluated over multiple plausible future meal scenarios, and the worst‐case violation is penalized.  A soft penalty on any predicted constraint breach steers the optimizer toward conservative dosing only when necessary, ensuring both efficacy and safety in the personalized treatment schedule.  

Once the MPC selects the optimal plan, only the first day’s recommendation is delivered to the patient (or care provider).  As new symptom reports arrive, the system slides its data window forward, updates the BNN input, and re–solves the MPC problem at the next dosing interval.  Optional clinician overrides or hard safety bounds (e.g.\ maximum daily dose) can be enforced at any stage.  Periodic retraining or online fine-tuning of the BNN ensures that the surrogate model remains calibrated to the patient’s evolving physiology or adherence patterns.

While in this paper we demonstrate our framework using a BNN, the architecture readily accommodates any uncertainty‐aware model (e.g.\ Gaussian processes or ensemble neural networks).  Moreover, the input dimension can be extended beyond simple symptom scores to incorporate additional noninvasive modalities—such as wearable pH sensors, activity trackers, or meal photographs—seamlessly within both the surrogate and MPC modules.  


\section{Gastric Acid Secretion and Patient Symptom Models}\label{sec:modeling}

\begin{figure}[thpb]
      \centering
      \includegraphics[width=0.75\linewidth]{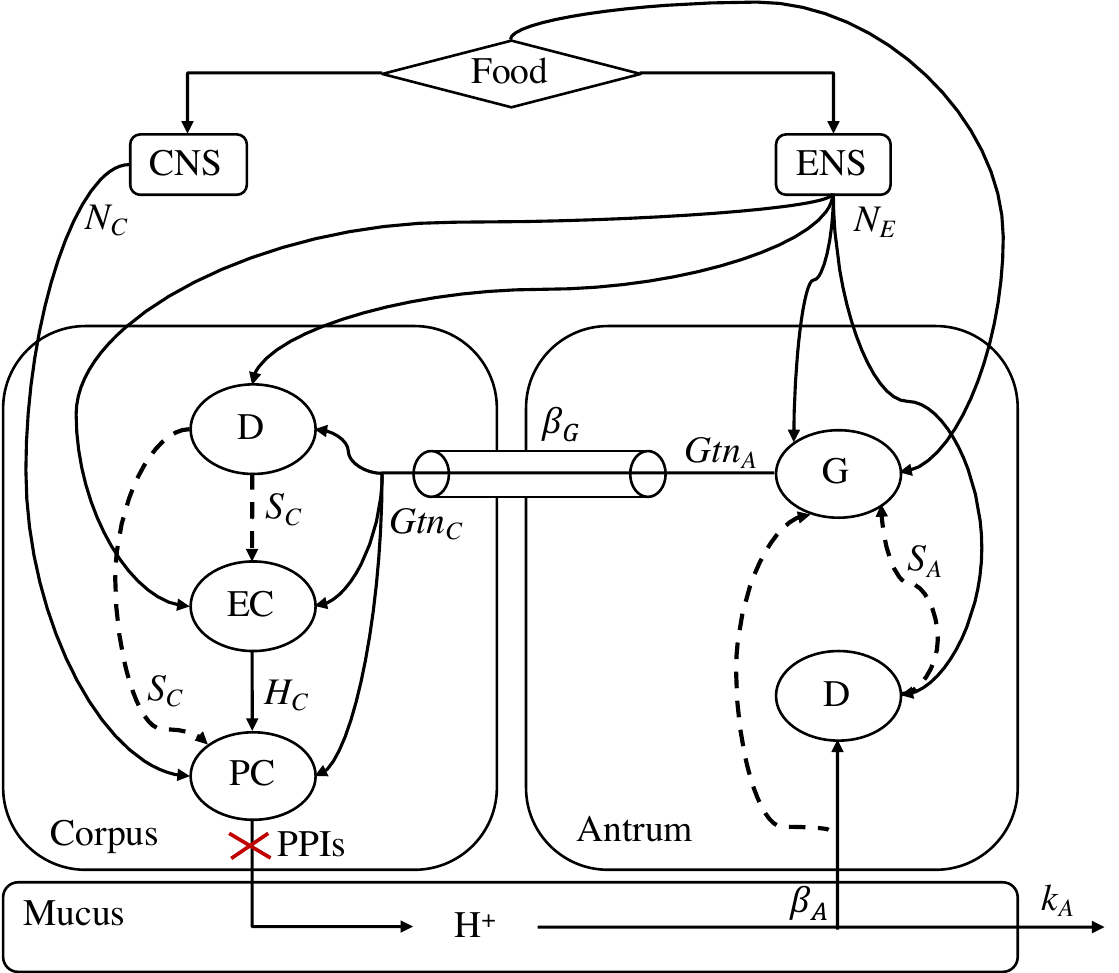}
      \caption{Schematic of gastric acid secretion regulation. Following food intake, central (CNS) and enteric (ENS) neural stimuli activate antral G cells to secrete gastrin ($Gtn_A$), which diffuses to the corpus to stimulate parietal cells (PC) to secrete acid ($H^+$) and to induce enterochromaffin‐like (ECL) cells to release histamine ($H_C$). Somatostatin from antral ($S_A$) and corpal ($S_C$) D cells inhibits these pathways. Solid arrows denote stimulation; dashed arrows denote inhibition. Transport of acid between compartments occurs at rate $\beta$, and luminal washout at rate $k_A$. The red cross indicates the site of PPI action on proton pumps.}
      \label{fig:GastricMdl}
\end{figure}

We employ the dynamic two‐compartment gastric acid secretion model with PPI inputs from \cite{sud2004predicting, joseph2003model, li2023scheduling} to both simulate true acid‐level trajectories and generate synthetic training data for our BNN.  As illustrated in Fig.~\ref{fig:GastricMdl}, the model divides the stomach into corpus and antrum compartments.  After a meal, Central Neural Stimuli (CNS) and Enteric Neural Stimuli (ENS) increase: In the antrum, ENS directly drives G cells to secrete gastrin, which diffuses to the corpus and activates parietal cells.  Simultaneously, gastrin and ENS stimulate enterochromaffin‐like (ECL) cells to release histamine, potentiating gastrin’s effect on acid secretion.  To maintain homeostasis, D cells in the corpus secrete somatostatin, which inhibits both gastrin and histamine pathways.  Finally, PPIs suppress acid output by blocking proton pumps on the parietal‐cell membrane.

This mechanistic framework captures the complex interplay of stimulatory and inhibitory pathways—and the pharmacodynamic impact of PPI dosing—necessary for generating high‐fidelity ground truth.  The full system of hormonal effector dynamics, acid and bicarbonate kinetics, and neural stimulus equations is detailed in our prior work \cite{li2023scheduling}.  Below, we introduce the newly added patient symptom models that map simulated acid concentrations to clinically relevant, noninvasive symptom scores.

To model the patient, we map the simulated gastric acid concentrations \(A_c(t)\) to two patient‐reported symptom scores—acid reflux and digestive discomfort—on a 1–10 scale.  Both scores use sigmoid mappings that reflect increasing symptom severity as acid levels deviate from healthy thresholds. We model the acid reflux score as

{\small\begin{equation}\label{eq:reflux_score}
S_{\mathrm{reflux}}(A_c)
= 1 \;+\;  \;\frac{9}{1 + \exp\bigl(-k_r\,(A_c - a_{\mathrm{high}})\bigr)},
\end{equation}}
where \(a_{\mathrm{high}}\) is the upper acid threshold above which reflux symptoms sharply increase, and \(k_r\) controls the sigmoid steepness. The digestive issue score is as follows,

{\small\begin{equation}\label{eq:digestive_score}
S_{\mathrm{digestion}}(A_c)
= 1 \;+\;  \;\frac{9}{1 + \exp\bigl(k_d\,(A_c - a_{\mathrm{low}})\bigr)},
\end{equation}}
where \(a_{\mathrm{low}}\) is the lower acid threshold below which digestive discomfort rises, and \(k_d\) sets the transition slope.

To mimic real‐world variability in patient reporting, we add Gaussian noise $\eta_r \sim \mathcal{N}(0,\sigma^2)$ and $\eta_d \sim \mathcal{N}(0,\sigma^2)$, and discretize to integer values:
{\small\begin{align}
\tilde S_{\mathrm{reflux}}(A_c)
&= \mathrm{clip}\Bigl(\,\lfloor S_{\mathrm{reflux}}(A_c) + \eta_r\rfloor,\;1,\,10\Bigr),\\
\tilde S_{\mathrm{digestion}}(A_c)
&= \mathrm{clip}\Bigl(\,\lfloor S_{\mathrm{digestion}}(A_c) + \eta_d\rfloor,\;1,\,10\Bigr),
\end{align}}

\noindent\text{where} 
\(a_{\mathrm{high}},a_{\mathrm{low}}\) are the symptom thresholds; 
\(k_r,k_d\) are sigmoid steepness parameters; 
\(\sigma\) is the noise standard deviation; 
\(\lfloor\cdot\rfloor\) rounds down to the nearest integer; 
\(\mathrm{clip}(x,1,10)\) bounds \(x\) within \([1,10]\) and $\eta_r$, $\eta_d$ vary with the patient but not with time. 

We model each patient’s meal intake \(Fd(t)\) as the sum of three Gaussian “meal pulses” per day—breakfast, lunch, and dinner—with randomized timing, amplitude, and width for $N_{\text{days}}$:

{\small\begin{equation}\label{eq:food_intake}
Fd(t)=\sum_{i=0}^{N_{\rm days}-1}\sum_{k=1}^{3}
A_{i,k}\,\exp\!\Bigl(-\tfrac{1}{2}\Bigl(\tfrac{t - (i + \mu_{i,k})}{\sigma_{i,k}}\Bigr)^{2}\Bigr).
\end{equation}}

Here, \(i\) indexes each full day and \(k\in\{1,2,3\}\) the three meals.  The pulse amplitude \(A_{i,k}\) is drawn from a uniform range (e.g.\ breakfast: \(\mathcal U(0.6,1.2)\), lunch: \(\mathcal U(0.6,1.5)\), dinner: \(\mathcal U(0.5,1.8)\)), the center time 
$\mu_{i,k}=C_k+\Delta_{i,k}$ with nominal meal times $C_1=8/24$, $C_2=12/24$, $C_3=18/24$ (days) and shift $\Delta_{i,k}\sim\mathcal U(-1.5/24,1.5/24)$, and the spread $\sigma_{i,k}\sim\mathcal U(0.2/24,1/24)$ (12–60 min).  Summing these three pulses each day produces a realistic, time‐varying food‐intake profile used both to train the BNN and to generate meal scenarios within the MPC.

\section{Bayesian Neural Network For Patient Symptom Prediction}
\label{sec:BNN}

\begin{figure}[t]
  \centering
  \includegraphics[width=0.9\linewidth]{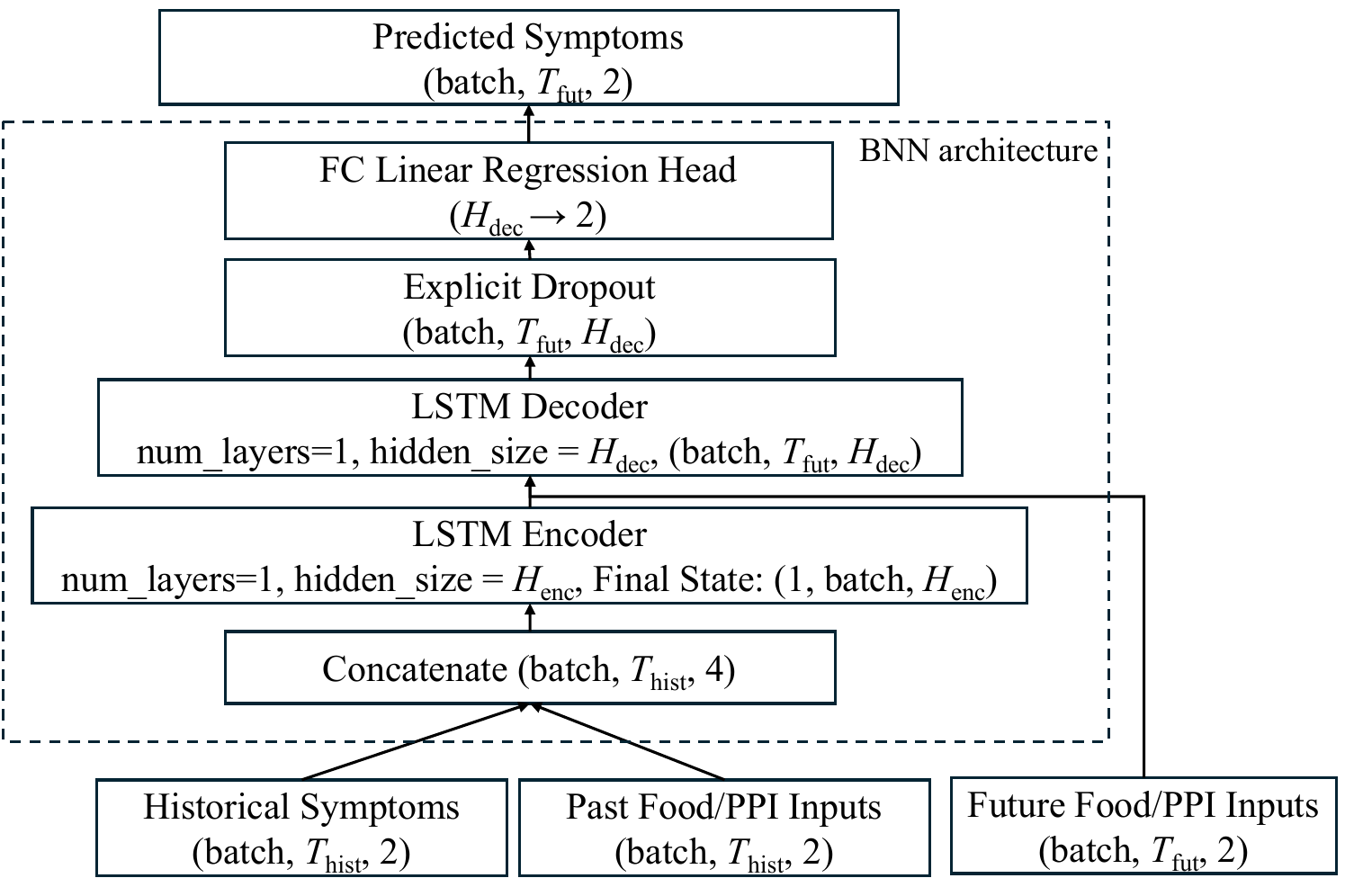}
  \caption{Sequence‐to‐sequence Bayesian Neural Network architecture.  The encoder LSTM processes the past $T_{\rm hist}$ timesteps of symptom scores and meal/PPI inputs into a hidden state, which initializes the decoder LSTM.  The decoder uses $T_{\rm fut}$ future meal/PPI inputs and, after Monte Carlo dropout, a linear regression head produces continuous reflux and digestive symptom predictions at each future timestep.}
  \label{fig:bnn_arch}
\end{figure}

In this section, we present the sequence-to-sequence Bayesian Neural Network (BNN), illustrated in Fig.~\ref{fig:bnn_arch}, which learns to predict future symptom trajectories from recent histories of reflux severity, digestive discomfort, meal intake, and PPI dosing. By leveraging Monte Carlo dropout, the BNN delivers both point estimates and principled uncertainty bounds. See \cite{gal2016dropout} for introduction to BNN; in the sequel, we use PyTorch \cite{paszke2019pytorch} for BNN training.  

As illustrated in Fig.~\ref{fig:bnn_arch}, at each control interval the BNN consumes two input tensors:  
First, a history tensor of shape \(\bigl[\mathrm{batch},\,T_{\rm hist},\,2\bigr]\) containing the past \(T_{\rm hist}\) values of reflux severity and digestive discomfort where batch stands for batch size. Second, a combined‐input tensor of shape \(\bigl[\mathrm{batch},\,T_{\rm hist}+T_{\rm fut},\,2\bigr]\) containing normalized meal intensity and PPI dose for both the historical period and the future prediction horizon \(T_{\rm fut}\).

These tensors are concatenated per timestep into 4-dimensional feature vectors and passed through a sequence-to-sequence LSTM architecture.  During the encoding phase, the encoder LSTM processes the \(T_{\rm hist}\) historical feature vectors one by one, producing a final hidden state that compactly summarizes the patient’s recent physiological state.  In the decoding phase, the decoder LSTM is initialized with encoded hidden state and uses only the \(T_{\rm fut}\) future meal/PPI inputs to generate a sequence of future hidden states that capture the expected symptom evolution.

To approximate a Bayesian posterior, fixed‐rate dropout is applied to each decoder hidden state before the regression head.  Finally, a linear layer maps each dropped-out hidden vector to two continuous outputs—predicted reflux severity and digestive discomfort—at each timestep in the prediction horizon.  

We note that framing symptom prediction as a regression task outperformed a classification‐based formulation. Although symptoms are reported on an ordinal 1–10 scale, treating them as discrete classes led to unstable training, coarse predictions at class boundaries, and lower overall accuracy.  By directly predicting continuous symptom values and optimizing a mean‐squared error loss, the model benefits from smoother gradients and can capture subtle temporal variations in patient reports.

For BNN training, we assemble a dataset by sliding a window of length \(T_{\rm hist}+T_{\rm fut}\) over long time‐series of (symptoms, meal, PPI) tuples.  The model is trained to minimize mean‐squared error between its \(T_{\rm fut}\)-step predictions and the true symptom values, using stochastic gradient descent with momentum and weight‐decay regularization.  Early stopping on a held‐out validation set was used to prevent overfitting.  To capture population‐level patterns, we first train a \emph{foundation} BNN on aggregated data from multiple patients (or synthetic simulations), then optionally fine‐tune on each patient’s own history for rapid personalization.

At runtime, we enable dropout at inference time and perform \(M\) stochastic forward passes through the trained BNN.  This Monte Carlo procedure yields \(M\) sample trajectories of predicted symptoms.  We summarize these into a per‐timestep mean \(\mu_t\) and standard deviation \(\sigma_t\), which serve as inputs to the chance‐constrained MPC.  

\section{Personalized PPI Dosage Scheduling via Chance‐Constrained MPC}
\label{sec:MPC}

In this section, we integrate the BNN‐based symptom predictor into a chance‐constrained Model Predictive Control (MPC) framework.  At each decision point, the MPC optimizer identifies the minimal sequence of daily PPI doses that ensures predicted symptom scores remain below a clinician‐defined threshold with high confidence (e.g., 95 \%).  To account for uncertainty in future meal intake, every candidate dosing schedule is evaluated across a set of plausible meal scenarios, and the worst‐case predicted constraint violation is penalized in the objective.

Let the prediction horizon be \(T\) timesteps, and let
{\small\[
u(t)\,,\quad t=1,\dots,T,\quad u(t)\in U\subset[0,1]
\]}
denote the normalized PPI dose at time \(t\).  We account for uncertainty in future meals by considering a finite set of \(J\) candidate disturbance profiles \(\{d^j\}_{j=1}^J\).  The BNN predictor
{\small\[
f\bigl(u(1\!:\!T),\,d(1\!:\!T)\bigr)
\;\mapsto\;
\bigl(\mu_i(t;u,d),\,\sigma_i(t;u,d)\bigr)_{i=1,2}^{t=1,\dots,T}
\]}
yields mean forecasts \(\mu_i\) and standard deviations \(\sigma_i\) for each symptom \(i\in\{1,2\}\) (reflux, digestive).

We aim at ensuring that each symptom remains below a clinician‐set threshold \(\theta\) with probability at least \(p\).  Approximating this chance constraint by averaging over the \(J\) meal scenarios gives:
{\small\[
\frac{1}{J}\sum_{j=1}^J
\mathbf{1}\{\mu_i(t;u,d^j)\leq\theta\}
\;\ge\;p
\quad\forall\,t=1,\dots,T,\;i=1,2.
\]}

Define the scenario‐averaged uncertainty
{\small\[
\hat\sigma_i(t;u)
\;=\;\frac{1}{J}\sum_{j=1}^J\sigma_i\bigl(t;u,d^j\bigr).
\]}
We then consider the following finite‐horizon dynamic optimization problem:
{\small\begin{subequations}\label{equ:ppi_mpc}
\begin{align}
\min_{\,u(1:T)\,\in\,U}\quad
&\sum_{t=1}^T c\,u(t)
\;+\;\lambda\sum_{t=1}^T\sum_{i=1}^2 \hat\sigma_i(t;u)\,,\label{eq:obj}\\
\text{s.t.}\quad
&\frac{1}{J}\sum_{j=1}^J
\mathbf{1}\{\mu_i(t;u,d^j)\leq\theta\}
\;\ge\;p,\quad \\& \forall\,t=1,\dots,T,\;i=1,2.\label{eq:chance}
\end{align}
\end{subequations}}
Here \(c\) penalizes total drug usage and \(\lambda\) penalizes residual uncertainty, yielding a dosing sequence that is both parsimonious and robust to meal variability.

The MPC problem in \eqref{equ:ppi_mpc} poses two significant challenges in practice:  
1) \emph{Recursive feasibility} is not guaranteed—once the horizon moves forward, there may be no admissible \(u(t)\) satisfying the chance‐constraints for all future steps.  
2) The chance‐constraint \eqref{eq:chance} is combinatorial and nonconvex, since it involves indicator functions and scenario‐averaging over \(J\) meal profiles.  

To address both issues simultaneously, we replace the hard probability constraints with a smooth penalty on the worst‐case violation and exploit a Gaussian approximation of the BNN forecasts.  Specifically, we assume each symptom forecast  
\((\mu_i(t;u,d^j),\sigma_i(t;u,d^j))\)  
follows a Normal distribution, so that the chance‐constraint  
\(\Pr\bigl(\text{symptom}_i(t)\!\leq\!\theta\bigr)\ge p\)  
is conservatively enforced by the linear bound  
\(\mu_i + \beta\,\sigma_i \le \theta\), where \(\beta = \Phi^{-1}(p)\) and $\Phi^{-1}$ is the quantile function (inverse CDF) of the standard normal distribution (e.g.\ \(\beta\approx1.28\) for \(p=0.9\)).  We then absorb any remaining violation into the objective via a rectified penalty:

{\small\begin{align}\label{equ:reformulated_mpc}
&\min_{\,u(1:T)\in U}\quad
\sum_{t=1}^T c\,u(t)
\;\\&\notag+\;\lambda \sum_{t=1}^T\;\sum_{i=1}^2
\max_{j=1,\dots,J}\Bigl[\mu_i\bigl(t;u,d^j\bigr)
+\beta\,\sigma_i\bigl(t;u,d^j\bigr)\;-\;\theta\Bigr]_+,
\end{align}}

\begin{algorithm}
\caption{Simulation‐Based MPC Dose Search}
\label{alg:mpc_search}
\begin{algorithmic}[1]
  \State \textbf{Input:} BNN predictor $\mathcal{F}$, symptom history $Y_{\rm hist}$,\\
  \hspace{2em}meal history $D_{\rm hist}$, PPI history $U_{\rm hist}$,\\
  \hspace{2em}meal forecasts $\{D^j\}_{j=1}^J$, candidates $\{U^k\}_{k=1}^K$,\\
  \hspace{2em}threshold $\theta$, penalty $\lambda$, confidence $\beta$
  \State \textbf{Output:} next‐interval dose $u_{\rm apply}$
  \State best\_score $\gets +\infty$
  \State best\_sequence $\gets []$
  \For{$k = 1$ \textbf{to} $K$}
    \State $U_{\rm cand} \gets U^k$
    \State violations $\gets []$
    \For{$j = 1$ \textbf{to} $J$}
      \State $(\mu,\sigma)\gets \mathcal{F}(Y_{\rm hist},D_{\rm hist},U_{\rm hist},U_{\rm cand},D^j)$
      \State $v\gets 0$
      \For{$t = 1$ \textbf{to} $T$}
        \State $\delta_{1}\gets \mu_{t,1} + \beta\,\sigma_{t,1} - \theta$
        \State $\delta_{2}\gets \mu_{t,2} + \beta\,\sigma_{t,2} - \theta$
        \State $v\gets v + \max(0,\delta_{1}) + \max(0,\delta_{2})$
      \EndFor
      \State violations.\texttt{push\_back}($v$)
    \EndFor
    \State $w\gets \max(\text{violations})$
    \State usage $\gets \sum_{t=1}^T U_{\rm cand}(t)$
    \State score $\gets$ usage $+$ $\lambda\,w$
    \If{score $<$ best\_score}
      \State best\_score $\gets$ score
      \State best\_sequence $\gets U_{\rm cand}$
    \EndIf
  \EndFor
  \State $u_{\rm apply}\gets \text{best\_sequence}[1]$
  \State \Return $u_{\rm apply}$
\end{algorithmic}
\end{algorithm}

We solve \eqref{equ:reformulated_mpc} using Algorithm \ref{alg:mpc_search}. 
We restrict PPI adjustments to three discrete actions—“increase by 20\%,” “maintain,” or “decrease by 20\%”—to avoid abrupt dosing changes.  The MPC search enumerates all length-$T$ sequences of these actions, uses the BNN to forecast mean and uncertainty of reflux and digestive symptoms under each sequence and each meal scenario, computes a rectified penalty for any predicted exceedance of the clinician’s threshold, and selects the sequence minimizing the sum of total PPI usage and the worst‐case violation.  Only the first day’s dose is applied; the procedure repeats at the next update time instant.  

\section{Simulations and Results}\label{sec:results}

To evaluate our framework, we conducted in silico experiments on a population of virtual patients whose gastric‐acid kinetics differ via randomly sampled model parameters. Each patient’s true acid dynamics and symptom trajectories are generated by the high‐fidelity ODE model from Section~\ref{sec:modeling}. 
For each virtual patient, we sample model parameters uniformly within physiologically plausible bounds to produce a unique gastric response. We simulate 390 days of data at an hourly resolution (24 samples/day), recording meal intensity, PPI dose, and the two symptom scores (reflux and digestion) as per Section~\ref{sec:modeling}. All signals share a common, hourly time base.
\begin{figure*}[t]
  \centering
  \includegraphics[width=0.95\textwidth]{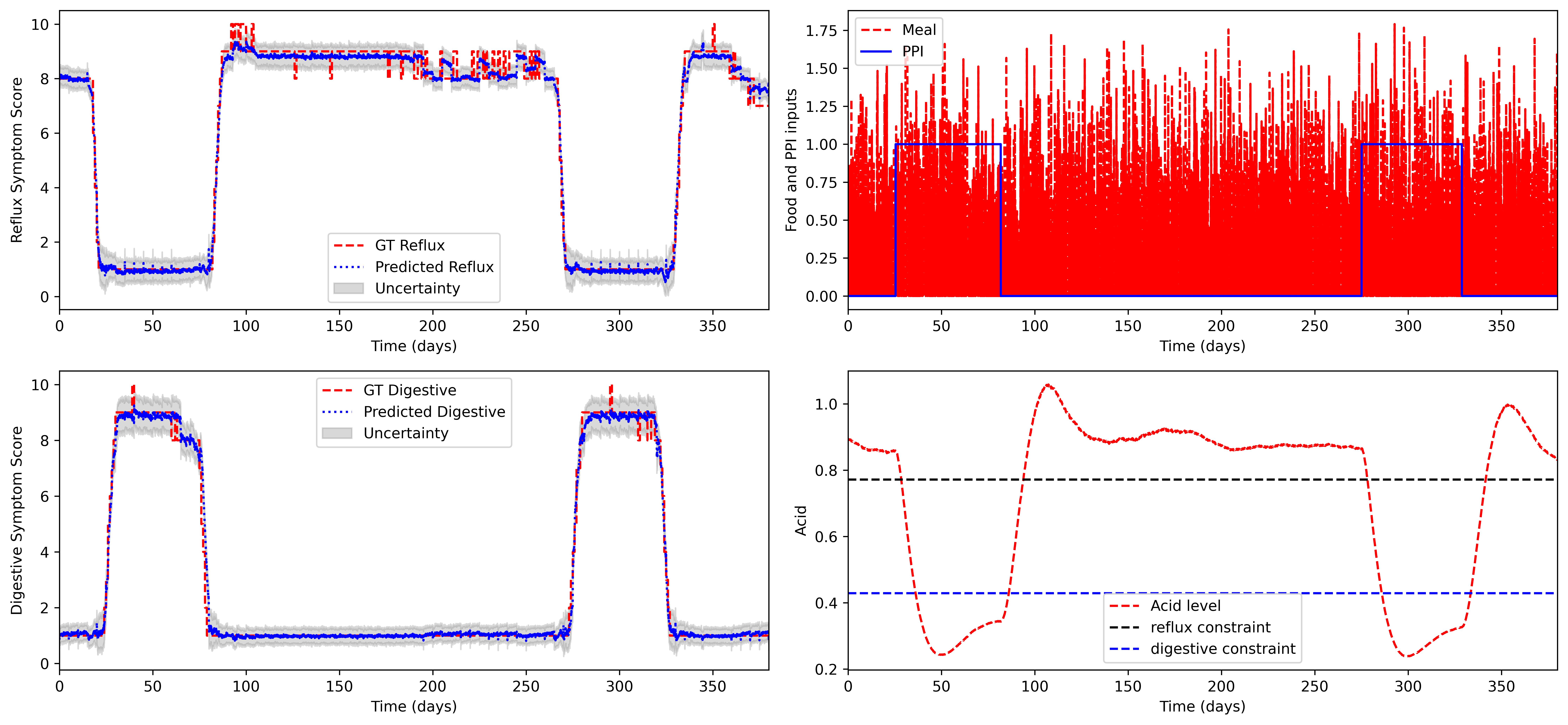}
  \caption{BNN open-loop validation over 300 days for a representative virtual patient.  (Top bottom left) True vs.\ predicted reflux and digestive scores (dotted red vs.\ solid blue), with MC‐dropout uncertainty (gray).  (Top bottom right) Underlying meal (red) and PPI (blue) inputs, and hidden acid level (red) with the symptom thresholds (dashed).  Across the population, the BNN achieves RMSE\(<0.5\) for both symptoms.}
  \label{fig:BNN_validation}
\end{figure*}
\begin{figure*}[t]
  \centering
  \includegraphics[width=0.95\textwidth]{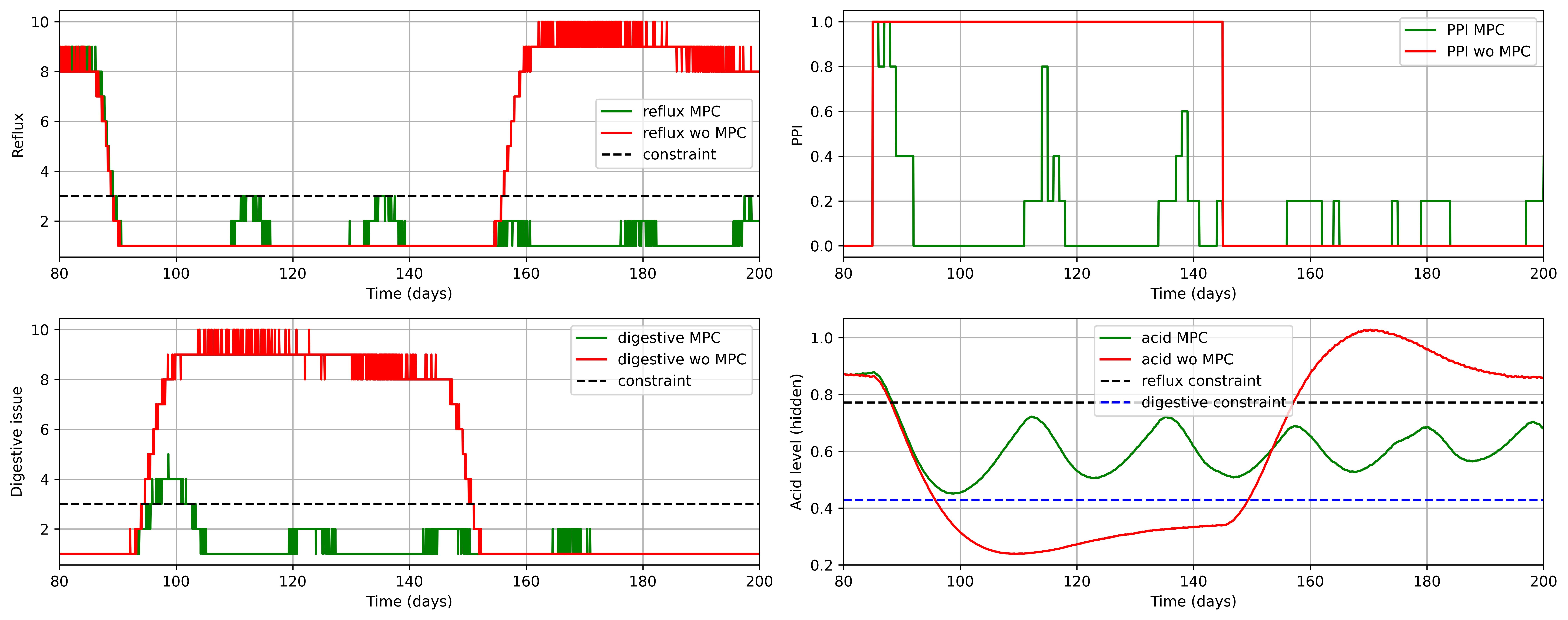}
  \caption{Closed-loop MPC vs.\ fixed dosing over days 80–200 for one patient.  Top bottom left panels: symptom trajectories under MPC (green) vs.\ without MPC (red), with threshold (black dashed).  Top bottom right: PPI dose profiles and hidden acid levels.  MPC enforces constraints and uses substantially less drug.}
  \label{fig:mpc_result}
\end{figure*}
We begin by training a foundation BNN on an aggregated dataset from 100 virtual patients.  For each patient, we simulate 60 days of randomized meal/PPI schedules and record the resulting symptom trajectories.  This yields an input tensor of size \((60 \times 24)\times 3\) (meal intensity, PPI dose, and past symptom scores) and a target tensor of size \((60 \times 24)\times 2\) (future symptom scores).  The sequence‐to‐sequence foundation BNN is then trained on this pooled dataset until both training and validation losses converge.  
Next, for each test patient we simulate 30 days of symptom history under randomized meal and PPIs patterns. With the encoder parameters held fixed, we then fine‐tune the foundation BNN’s decoder LSTM and final regression head on this patient‐specific data, rapidly adapting the model to individual gastric dynamics.

Figure~\ref{fig:BNN_validation} shows one representative patient simulation results. During the simulation, we freeze the BNN weights and run the model open‐loop for 300 days of new, randomly generated meal schedules. The BNN’s mean prediction (blue line) tracks the ground truth (red dashes) with a root‐mean‐square error below 0.5 score‐units in both channels, and uncertainty bands (gray) appropriately widen during abrupt symptom changes.

Finally, we integrate the fine‐tuned BNN into the MPC framework and simulate 300 days of closed‐loop, personalized PPI dosing for each virtual patient.  Meal intake is sampled hourly from randomized profiles, and for each 24-hour interval the MPC solver (Algorithm~\ref{alg:mpc_search}) selects one of three dose adjustments (\(-20\%\), \(0\%\), or \(+20\%\)).  We benchmark against a conventional twice-daily fixed regimen set at the 95th-percentile open-loop dosing requirement.  Over 20 virtual patients, our symptom-driven, learning-augmented MPC maintains both reflux and digestive scores below clinician-defined thresholds at least 95\% of the time, while reducing total PPI consumption by over 65\% compared to the fixed schedule. Moreover, by dynamically adapting doses, our approach mitigates rebound hyperacidity—administering smaller, more gradual PPI adjustments to suppress post-treatment acid spikes—without any invasive pH monitoring.

\section{Conclusion}
We have presented a fully noninvasive, symptom‐driven framework for personalized PPI therapy that relies exclusively on patient‐reported reflux and digestive discomfort.  A Bayesian Neural Network model learns to predict future symptom trajectories—and to quantify its own uncertainty—using only historical symptom scores, meal logs, and dose records. These probabilistic predictions are embedded into a chance‐constrained Model Predictive Control scheme, which computes the minimal daily PPI dosage required to suppress symptoms with high confidence, all without any invasive gastric acid monitoring.  In silico experiments across varied dietary schedules and virtual patient populations demonstrate that our learning‐augmented MPC achieves over 65\% reduction in total PPI usage compared to standard fixed regimens, while maintaining symptom control with at least 95\% reliability. This hybrid approach paves the way for practical, patient‐centric PPI dosing that minimizes both treatment burden and overdose risk.



\bibliography{references.bib}{}
\bibliographystyle{IEEEtran} 

\end{document}